# Interpretable QSPR Modeling using Recursive Feature Machines and Multi-scale Fingerprints

Jiaxuan Shen[1], Haitao Zhang[1], Yunjie Wang[1], Yilong Wang[1], Song Tao[1], Bo Qiu[1,2]*, Ng Shyh-Chang[3,4,5,6]*

[1]Hebei University of Technology, Tianjin, China

[2]University of Science and Technology Beijing, Beijing, China

[3]Institute of Zoology, Chinese Academy of Sciences, Beijing, China

[4]Institute for Stem Cell and Regeneration, Chinese Academy of Sciences, Beijing, China

[5]Beijing Institute for Stem Cell and Regenerative Medicine, Beijing, China

[6]University of Chinese Academy of Sciences, Beijing, China

*Corresponding authors: huangsq@ioz.ac.cn (N. Shyh-Chang); 2404565@ustb.edu.cn (B. Qiu)

## Abstract

Interpretable machine learning is vital for scientific research, especially in biology and chemistry, where it facilitates advances in drug discovery and molecular design. Quantitative Structure-Property Relationship (QSPR) models have made significant strides in drug development, but many existing QSPR models lack interpretability, which hinders the development of new drugs and molecules. This study pioneers the application of Recursive Feature Machines (RFM) in QSPR modeling, introducing a tailored feature importance analysis approach to enhance interpretability. By leveraging deep feature learning through AGOP, RFM achieves state-of-the-art (SOTA) results in predicting molecular properties, as demonstrated through solubility prediction across nine benchmark datasets. To capture a wide array of structural information, we employ diverse molecular representations, including MACCS keys, Morgan fingerprints, and a custom multi-scale hybrid fingerprint (HF) derived from global descriptors and SMILES local fragmentation techniques. Notably, the HF offers significant advantages over MACCS and Morgan fingerprints in revealing structural determinants of molecular properties. The feature importance analysis in RFM provides robust local and global explanations, effectively identifying structural features that drive molecular behavior and offering valuable insights for drug development. Additionally, RFM demonstrates strong redundancy-filtering abilities, as model performance remains stable even after removing redundant features within custom fingerprints. Importantly, RFM introduces the deep feature learning capabilities of the average gradient outer product (AGOP) matrix into ultra-fast kernel machine learning, to imbue kernel machines with interpretable deep feature learning capabilities. We extend this approach beyond the Laplace Kernel to the Matern, Rational Quadratic, and Gaussian kernels, to find that the Matern and Laplace kernels deliver the best performance, thus reinforcing the flexibility and effectiveness of AGOP in RFM. Experimental results show that RFM-HF surpasses both traditional machine learning models and advanced graph neural networks, excelling in both predictive accuracy and interpretability. In summary, this study establishes RFM as a powerful and interpretable tool in molecular property prediction, with the proposed feature importance analysis providing critical guidance for molecular design and drug discovery.

**Keywords**: machine learning; average gradient out product; recursive features machine; feature importances analysis; molecular solubility; multi-scale hybrid fingerprints

# 1 Introduction

In the fields of chemistry and pharmaceuticals, certain molecular properties are key factors influencing drug development and application. Accurately predicting these properties is crucial for molecular design, accelerating new drug development, optimizing drug formulations, and improving drug bioavailability. Although traditional experimental methods can provide precise data, they are often time-consuming and costly, limiting research efficiency. With the rapid advancement of computer science and technology, significant progress has been made in recent years in predicting molecular properties using machine learning (ML) methods[1]-[6].

As a powerful computational tool in the drug discovery process, QSPR models play a vital role in virtual screening and lead compound optimization stages. Experimentally testing the properties of candidate compounds typically requires substantial resources and time, whereas QSPR models offer a relatively convenient in silico alternative. These models predict the quantitative relationship between molecular structure and physicochemical properties, providing valuable feedback to biochemists in experimental design and prioritization. Consequently, QSPR models have become an efficient auxiliary tool in drug development, helping researchers conduct more targeted experimental studies, thereby improving research efficiency. QSPR models are based on the hypothesis that a molecule's structure determines its properties, and by analyzing and quantifying the structural features of molecules, their physicochemical properties can be predicted. The development of QSPR models generally involves two steps: first, the description and quantification of molecular structures, and second, the establishment of a mathematical relationship between structure and properties[7]-[9]. The molecular representation used to build the model is crucial for accurately predicting molecular properties and biological activity. Classical molecular descriptors[10], molecular fingerprints[11], graphs[12], and molecular SMILES strings[1] are common and effective molecular representation methods.

After obtaining molecular representations, it is necessary to choose an effective modeling method to perform supervised modeling of the relationship between chemical representations and molecular properties. Common modeling methods include traditional machine learning algorithms, such as Support Vector Machines (SVM)[13]-[14], Random Forests (RF)[15], Kernel Ridge Regression (KRR)[16], k-Nearest Neighbors (KNN)[17], and deep learning methods based on deep neural networks (DNNs), such as Multilayer Perceptrons (MLP)[18], Convolutional Neural Networks (CNN)[19], Recurrent Neural Networks (RNN)[20], and Graph Neural Networks (GNN)[21]-[22]. The choice of modeling approach often depends on the type of molecular representation used. For example, when using graphs as molecular representations, Message Passing Graph Neural Networks (MP-GNN)[21] can effectively integrate information from nodes and edges in the molecular graph through the message-passing mechanism, thereby capturing complex structural and topological relationships.

Although some models exhibit excellent predictive performance, most models still suffer from poor interpretability, especially deep learning QSPR models based on neural networks. As computational power increases, the "black-box effect" of neural network models becomes more pronounced, with issues such as a large number of parameters with unclear roles, nonlinear transformations in unknowably high-dimensional spaces, and the opacity of the training optimization and feature learning process. However, to realize the goals of "AI for Science", model interpretability is particularly important. Thus ML researchers have proposed game-theoretic techniques such as LIME[23] and SHAP[24]-[25] to approximate local or global explanations for the decision processes of neural network models. To improve our understanding of the relationship between molecular structure and molecular properties, ML model interpretability is especially critical.

Recently, it was discovered that Recursive Feature Machines (RFM) possess the capacity for ultra-fast approximation and quantification of feature importance during deep feature learning by all DNNs[26]-[27]. By applying these RFMs to rapidly learn solubility datasets, then mapping the trained samples back to the trained AGOP matrix, we were able to score the importance of features for each sample, and the effectiveness of this scoring was validated through in silico experiments. This mapping of individual samples provides local interpretability, allowing us to quantify the positive or negative influence of features on molecular properties. Furthermore, by analyzing the local impacts across all samples, we can obtain global interpretability of the model and identify the dataset features that most significantly affect molecular properties. After comparing our method with existing popular interpretability models, we demonstrated the effectiveness of the proposed feature importance analysis method. In addition to feature importance analysis in molecular datasets, we discovered that RFM has the ability to filter redundant features. During the learning of custom multi-scale hybrid fingerprints (HF) that incorporate SMILES fragments, even when feature fragments with similar meanings were removed, the model's performance did not show any significant decline.

For molecular representation, we tested molecular MACCS-Keys fingerprints(167 bits)[28], Morgan fingerprints(2048bits)[29], and HF fingerprints to demonstrate the impact of different molecular representations on predictive performance. We compared the performance of RFM with other ML models across nine solubility datasets. The results showed that RFM-HF achieved superior performance against current SOTA DNNs and GNNs, and we also provided global feature importance scores for all substructures in these datasets, offering a valuable reference for molecular screening and design. Finally, we applied AGOP to other kernel functions besides the Laplace kernel and analyzed different data features, which led us to some observations on the overfitting phenomenon in the kernel ridge regression of RFMs.

## 2 Related Work

Model interpretability is generally divided into three categories: feature importance scores, identification of training instances with the greatest impact on predictions, and the generation of counterfactuals. Among these, research on feature importance scores is the most extensive. For example, there are pre-model statistical explanation methods, such as p-values[30], False Discovery Rate (FDR)[31], and Family Wise Error Rate (FWER)[32], as well as post-model explanation methods, like Permutation Importance (PI)[33], which determines feature importance by randomly shuffling feature values and observing the extent of performance degradation. Shapley value-based Additive Explanations (SHAP)[24-25] explains "black-box models" by casting the model in game theory and calculating the marginal contribution of each input feature or player to the model output or game score, providing both global and local interpretations. In addition to feature importance analysis methods for traditional ML, research on the interpretability of DNNs has also been developing rapidly. For example, it is now possible to calculate atomic, bond, and global importance scores using integrated gradients and achieve high model interpretability through an embedded attention mechanism[1].

Here, we are the first to apply Recursive Feature Machines (RFM)[26]-[27] to molecular QSPR modeling and remap the samples back to the trained feature matrix to perform feature importance analysis. At the same time, RFM demonstrated outstanding performance in QSPR modeling among many ML models. Considering that MACCS-Keys struggle with complex or uncommon substructures corresponding to molecular properties, and that Morgan fingerprints have issues with hash collisions, we customized a new set of multi-scale hybrid fingerprints (HF). Our RFM-HF outperformed existing fingerprints, ML methods and GNNs in predictive performance, and since the fingerprint bits correspond one-to-one with molecular substructures, it facilitates deeper analysis of the impact of various functional groups on solubility, and the derivation of new biochemical principles.

## 3 Methods

This chapter first introduces the representation methods used for the molecules, followed by an explanation of the principles of AGOP, RFM, and their feature importance analysis. It then provides an overview of kernel methods and the fundamental concepts of overfitting in these methods.

**3.1 Molecular Representation**

Molecular descriptors are mathematical representations used to quantify molecular structural features, encompassing various types such as topological, geometrical, electrical, and thermodynamic properties, which comprehensively reflect the structural information of molecules. Compared to molecular descriptors, molecular fingerprints provide a more dynamic representation by capturing the characteristics of molecules through fragment-based features. Molecular fingerprints are a binary-encoded molecular representation method that maps the chemical structure information of molecules into a fixed-length bit string, where the position of each bit indicates the presence or absence of a specific substructure or feature. Due to their high information density and computational efficiency, molecular fingerprints are widely used in rapid compound database searches and similarity analyses. Common molecular fingerprint algorithms include MACCS keys and ECFPs (Extended-Connectivity Fingerprints)[34].

We utilized a total of 9 datasets, including AqSolDB[35], Arash[36], ESOL[37], FreeSolv[38], and five subsets of Samuel[39]. Regarding molecular representation methods, MACCS Keys, 2048-bit Morgan fingerprints with a diameter of 2, and a custom multi-scale hybrid fingerprint composed of descriptors and SMILES fragmentation fingerprints was utilized. The predictive performance of various machine learning models under different representations was compared, revealing that the choice of representation has a significant impact on predictive performance, regardless of ML methods and neural network-based deep learning models.

**3.2 AGOP, KRR, RFM and Feature Importance Analysis**

This subsection is divided into three parts, introducing the basic principles of Average Gradient Outer Product (AGOP)[40]-[41], Kernel Ridge Regression (KRR)[42], and Recursive Feature Machines (RFM)[26]-[27] and how we implement feature importance analysis.

### 3.2.1 Basic Principles of AGOP

In recent years, DNNs have achieved remarkable success across various fields and scientific tasks, including RNN-based and Transformer-based language models, and CNN-based visual recognition models. The exceptional performance of DNN can be attributed to their ability to learn features from the data manifold. Understanding how DNNs learn features from data manifolds is crucial for our understanding of these ML models, enhancing their transparency, and improving their performance.

Consider a fully connected network with a depth L > 1. According to the universal approximation theorem, this network approximates an $f(x): \mathbb{R}^d \rightarrow \mathbb{R}$, as shown in Equation (1). The weight matrix $W^{(l)} \in \mathbb{R}^{k_{l+1} \times k_l}$, where $k_l$ is the hidden dimension of the $l$th layer. The nonlinear activation function is denoted as $\phi(x)$. The input of the $l$th layer ($l \in \{0, \dots, L-1\}$) is denoted as $x_l$, where $x_0 \equiv x$ represents the original data point.

Set $k_{L+1}$ to the number of output classes, and $k_0 = d$ as the input data dimension. In supervised learning, the mapping $f$ from input data to labels is learned by minimizing the loss function $\mathcal{L}(\theta, X)$ over input-label pairs, where $X$ is the input dataset and $\theta$ is the set of weights. Recent work defined the relevant structure captured by DNNs, as a Feature Matrix [26]-[27], which, for a given layer $l$, is defined as the Gram matrix of the columns of the weight matrix $W^{(l)}$, denoted as $F_l$, as shown in Equation (1).

$$F_l \equiv (W^{(l)})^T W^{(l)} \tag{1}$$

Using the Neural Feature Ansatz (NFA)[26]-[27], as shown in Equation (2), it was shown that the Neural Feature Matrix (NFM) of any intermediate layer of the trained network $f$ is always correlated to the Average Gradient Outer Product (AGOP), denoted as $\bar{G}_l$, where $\frac{\partial f(x)}{\partial x_l} \in \mathbb{R}^{k_l \times 1}$ represents the gradient of the function $f$ with respect to the intermediate representation $x_l$. For each layer $l$, the NFM and AGOP will have high similarity or correlation. The definition of correlation is provided in Equation (3), where A denotes NFM and B denotes AGOP. To improve computational efficiency, the gradients can be concatenated into a matrix $\frac{\partial f(X)}{\partial x_l} \in \mathbb{R}^{n \times k_l}$. When considering only regression tasks, the output dimension is 1. When the output dimension d>1, the relationship defined by the NFA remains the same, where $\frac{\partial f(x)}{\partial x_l} \in \mathbb{R}^{k_l \times d}$ represents the Jacobian matrix of the input-output relationship of the model $f$, which is a natural extension of the gradient and Laplacian.

$$(W^{(l)})^T W^{(l)} \propto \frac{1}{n} \sum_{\alpha}^{n} \frac{\partial f(x_l^{(\alpha)})}{\partial x_l} \frac{\partial f(x_l^{(\alpha)})}{\partial x_l}^T \equiv \bar{G}_l \qquad (2)$$

$$\rho(A,B) = tr(A^T B) \cdot tr(A^T A)^{-1/2} \cdot tr(B^T B)^{-1/2} \qquad (3)$$

For any layer $l$, the AGOP can be rewritten in the form of Equation (4). If $K^{(l)}$ is an identity matrix, this alignment is simple and holds precisely. Even when $K^{(l)}$ is non-trivial, empirical evidence suggests that their correlation remains high. The NFA connects DNNs feature learning with the mathematical operator of AGOP. Experimental results indicate that AGOP accurately captures the features learned within the DNNs architecture, including in both visual and language tasks.

$$\bar{G}_l = (W^{(l)})^T K^{(l)} W^{(l)} \qquad (4)$$

3.2.2 Principles of KRR

The emergence of the Neural Tangent Kernel (NTK) established a theoretical connection between infinitely wide neural networks and kernel learning methods, suggesting that neural networks, in the infinite-width limit, behave similarly to kernel machines like Kernel Ridge Regression (KRR)[26][27]. Despite this connection, the success of DNNs is attributed to their ability to learn hierarchical representations through feature learning, which allows them to capture complex patterns in the data. This feature learning mechanism is absent in traditional kernel methods, where the features are implicitly defined by the kernel and remain fixed throughout training. This distinction highlights a key difference between DNNs and kernel methods: the former learns representations adaptively, while the latter relies on a predefined feature map derived from the kernel function.

To formalize the general framework of kernel methods, consider a set of training points $X = \{x_1, ..., x_n\} \subseteq \mathcal{X}$, sampled independently and identically distributed (i.i.d.) from a distribution $\mu$. Let $f^* \in L^2_\mu(\mathcal{X})$ represent the target function, and the response variable be given by $y_i = f^*(x_i) + \epsilon_i$, where $\epsilon_i$ is noise that is i.i.d with zero mean and variance $\sigma^2$. KRR minimizes a regularized loss function to learn a predictor $f \in \mathcal{H}$, where $\mathcal{H}$ is the Reproducing Kernel Hilbert Space (RKHS) associated with the chosen function kernel $K$. RKHS is a space of functions where the inner product is defined by the kernel, allowing us to generalize linear methods to nonlinear data. The objective of KRR is to minimize the following expression:

$$\min_{f \in \mathcal{H}} \frac{1}{n}\sum_{i=1}^{n}(f(x_i) - y_i)^2 + \gamma_n \|f\|_{\mathcal{H}}^2 \tag{5}$$

where $\gamma_n > 0$ is the regularization parameter that controls the complexity of the model and prevent overfitting. The term $\|f\|_{\mathcal{H}}^2$ represents the norm of the function $f$ in the RKHS, enforcing smoothness in the learned function. Let $y = (y_1, \ldots, y_n)^T$ be the vector of respone variables. The solution to this optimization problem can be expressed as:

$$\hat{f}(x) = <\hat{\theta}(y), \phi(x)> \tag{6}$$

where

$$\hat{\theta}(y) = \phi(X)^T (K + n\gamma_n I)^{-1} y \tag{7}$$

and $K = K(X, X') = (K(x_i, x_j))_{i,j}^n$ is the kernel matrix. Here, $\phi(X) = [\phi(x_1), \ldots, \phi(x_n)]^T \in \mathbb{R}^{n \times p}$ represents the feature map derived from the kernel function, transforming the input data into a high-dimensional space where linear methods can be applied[42].

### 3.2.3 RFM and Feature Importance Analysis

The ability of the function's AGOP or Jacobian or Laplacian matrix to capture features learned by DNNs offers new opportunities for improving kernel machines in KRR. Empirical evidence has shown that such matrices can capture the features learned by various DNNs architectures, such as Transformer language models, CNNs, and MLPs[26-27]. Let the matrix $M$ represent the AGOP of a kernel machine that is undergoing training. The RFM enhances the model's predictive performance and feature interpretability by recursively optimizing the feature weighting matrix M. Its core is based on the construction of the Laplace kernel function, which measures the similarity between samples. The Laplace kernel matrix with the introduction of the matrix M is calculated as:

$$K(x, x') = \exp\left(-\gamma \cdot \sqrt{(x - x')^T M (x - x')}\right) \tag{8}$$

where $\gamma = 1/L$ is the inverse of the bandwidth parameter, and M is the feature weighting matrix. This matrix M is recursively optimized to capture the relative importance of different features, allowing the kernel function to more accurately reflect the relationships between samples in the feature space. In each iteration, the model coefficients $\alpha$ are first solved through kernel regression:

$$(K_{train} + \lambda I)\alpha = y_{train} \tag{9}$$

where $K_{train}$ is the kernel matrix of the training samples, and $\lambda$ is the regularization parameter. Then, the matrix M is updated using the AGOP to better represent the importance of features. After multiple iterations of optimization, the matrix M is used to compute the kernel matrix $K_{test}$ for the test data, and predictions are made using $\alpha$.

In terms of feature importance scoring analysis, after training, the data samples $X$ and the AGOP of the trained kernel machine need to be reinterpreted. By mapping the features from the trained matrix M, the model can compute local interpretability scores for individual sample features and aggregate information from multiple samples to obtain global interpretability. Specifically, each sample $x_i$ in the sample matrix $X_{train}$ is standardized, and then the feature score is calculated by a double inner product with the trained feature weighting matrix M. The above principle is shown in Equation (10) and the schematic diagram is shown in Figure 1.

$$score(x_i) = x_i^T M x_i \tag{10}$$

This formula reflects the contribution of each feature in an individual sample to the target variable, providing local feature interpretability. To obtain global interpretability, the scores of all samples are summed and averaged to derive the importance contribution of each feature across the entire dataset:

$$average\ score = \frac{1}{n}\sum_{i=1}^{n} score(x_i) \tag{11}$$

By taking the absolute value of the average scores and ranking the features, the most important features for the model can be identified. This method effectively combines both local and global interpretability analyses, enabling the RFM to explain the contribution of individual sample features while revealing the key features of the overall model.

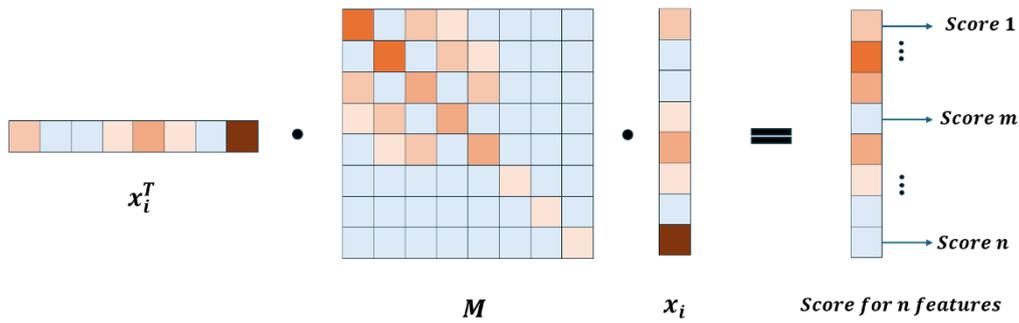

Figure 1 : RFM feature importance analysis schematic

**3.3 Demonstrating the Effectiveness of Feature Importance Analysis in RFM**

To validate the interpretability of the RFM model, we applied other popular feature importance analysis methods to the trained RFM model and performed a correlation analysis between the results of these methods and the directly computed RFM results. Permutation Importance (PI) and SHAP can be applied to any machine learning model without relying on the model's internal structure, making them widely used as general interpretability tools. These two methods evaluate feature importance by analyzing the impact of features on the model's output and attempt to quantify the contribution of each feature to the prediction result. To demonstrate the effectiveness of the RFM model in screening feature importance, we applied PI and SHAP to the trained RFM model and compared their analysis results with the directly computed RFM results, finding a high degree of correlation between them. Moreover, the computation time of the RFM was significantly lower than that of PI and SHAP.

In this study, we define correlation as follows: the feature importance scores of RFM, PI, and SHAP are calculated separately, and then ranked in descending order of importance. The top 10 or 20 most important features for each method are then selected, and the overlap of these features between RFM and PI, SHAP is compared. The more overlapping features there are, the stronger the correlation. The following sections describe the feature importance methods used.

(1) Permutation Importance

Permutation Importance is a model interpretation technique used to assess the contribution of each feature to the model's predictive performance. The core idea is to measure the importance of a feature by shuffling (permuting) its values and observing the change in the model's performance. Let the original feature matrix be $X$, the target variable be $y$, the model be $f$, and the performance metric be $M$ (e.g., accuracy, mean squared error). First, calculate the baseline performance $M_{baseline} = M(y, f(X))$. For each feature $X_j$, generate a permuted feature matrix $X_{permj} = (X_1, X_2, ..., shuffle(X_j), ..., X_p)$ by shuffling the values of $X_j$, and then calculate the performance after permutation $M_{permj} = M(y, f(X_{permj}))$. The importance of feature $j$ is defined as $I_j = M_{baseline} - M_{permj}$.

(2) SHapley Additive exPlanations (SHAP)

SHAP is based on the concept of Shapley values from game theory, assigning a contribution value to each feature, representing its influence on the model's prediction. SHAP ensures consistency and local accuracy and can provide explanations for various types of models. Let the model be $f$, and the feature vector be $x = (x_1, x_2, ..., x_n)$. The Shapley value $\phi_i$ for feature $i$ is defined as:

$$\phi_i = \sum_{S \subseteq N \setminus \{i\}} \frac{|S|!(n-|S|-1)!}{n!} [f(S \cup \{i\}) - f(S)] \quad (12)$$

where $N$ is the set of all features, $S \subseteq N\{i\}$ represents a subset of features excluding feature $i$, $|S|$ is the size of subset $S$, and $f(S)$ is the model output using only the features in subset $S$. Finally, the importance of the feature is determined based on the magnitude of its SHAP value.

### 3.4 Analysis of Overfitting Degrees in Kernel Functions

RFM represent an innovative approach that integrates the feature extraction mechanisms of neural networks into kernel learning. The original RFM, which incorporated the AGOP into the Laplace kernel, exhibited remarkable performance. To further explore the versatility of this method, we applied AGOP to additional kernel functions, specifically the Matern kernel, Gaussian kernel, and Rational Quadratic kernel. These four kernels exhibited varying degrees of overfitting during our experiments. Despite applying consistent regularization across all kernels, differences in generalization performance were still observed. These variations can be characterized by analyzing the spectral properties of the kernel matrices or the M matrices. Recent studies suggest that benign overfitting occurs when a model interpolates noisy training data with minimal impact on test performance[44]. On the other hand, catastrophic overfitting arises when the model's interpolation leads to unbounded test error[44]. Benign overfitting is closely linked to the spectral decay rate of the kernel matrix. It occurs only when the spectral decay is slower than any polynomial decay, specifically when $\lambda_k = \Theta(k^{-1-\epsilon})$ for some constant $\epsilon > 0$, such as in linear polylogarithmic decay where $\lambda_k = \Theta(k^{-1}\log^{-a}(k))$ with $a > 0$ [44]. However, such slow spectral decay is not typically observed in standard kernel methods.

By examining the performance of these four kernels on both the training and test sets, as well as analyzing the spectral properties of their post-training M matrices, we can assess their degrees of overfitting on the datasets.

# 4 Experiments

The experimental section begins with a detailed analysis of the datasets, covering sample size, molecular weight distribution, target value distribution, and the specific representation methods used for each dataset. This is followed by an overview of the experimental workflow. Finally, the experimental results are presented for readers' reference.

**4.1 Datasets**

Water and various organic solvents are indispensable in the fields of chemistry and biology, with solubility playing a critical role in processes such as extraction, metabolism, and protein engineering. Therefore, the measurement and prediction of solubility have always been of significant interest. Although laboratory measurements provide accurate data, the process is cumbersome and time-consuming, making data-driven predictive models of great importance.

For our dataset selection, we used the AqSolDB aqueous solubility dataset, Samuel's datasets for aqueous solubility and organic solvents (comprising five subsets), Arash's dataset for the solubility of organic compounds in water, as well as the ESOL and FreeSolv datasets. Water-soluble compounds can be classified based on their solubility values (LogS): compounds with solubility ≥ 0 are considered highly soluble, those with solubility between 0 and -2 are soluble, between -2 and -4 are sparingly soluble, and those with solubility below -4 are deemed insoluble.

Due to differences in experimental conditions, measurement methods, and units across existing datasets, AqSolDB standardized and compiled nine datasets, ultimately containing 9,982 unique compounds. The AqSolDB data is stored in CSV file format and includes the SMILES representation of the compounds, and experimental aqueous solubility data. The data used by Samuel et al. was sourced from the Open Notebook Science Challenge's aqueous solubility dataset and the Reaxys database, collecting solubility data for water, ethanol, benzene, and acetone. Arash's dataset selected Vermeire's (11,804 data points), Boobier's (901 data points), and Delaney's (1,145 data points) data, ultimately forming a dataset of 8,438 organic compounds and their solubility data by omitting non-unique measurements and noisy data. The ESOL dataset contains experimental solubility data for 1,128 organic molecules, while the FreeSolv dataset provides solvation free energy data for 642 molecules in water. These datasets offer a broader scope for solubility and hydration studies, helping to enrich the diversity of molecular representations and physical properties used in our analysis.

AqSolDB and Arash are the largest datasets with nearly 10,000 and 8,500 data points, respectively. FreeSolv and the Samuel datasets are considerably smaller, ranging from a few hundred to just over 1,000. This difference in size can significantly impact model training and performance. Aqueous datasets (AqSolDB, ESOL) generally exhibit lower (more negative) mean and median solubility values, indicating that the molecules in these datasets tend to have lower water solubility on average. The Samuel datasets, especially Samuel_water_narrow and Samuel_benzene, exhibit smaller standard deviations in molecular weight compared to other datasets. This suggests a more homogenous set of molecules within each of these smaller datasets. The datasets exhibit considerable variability in both size and molecular properties. This variability is crucial to consider when choosing datasets for model training and evaluation. Larger datasets like AqSolDB and Arash offer more data points but also greater diversity in both solubility and molecular weight. Smaller datasets like FreeSolv and the Samuel sets offer more focused subsets of molecules, but their smaller size and potential biases should be taken into account. Understanding these characteristics is vital for selecting the right data for building robust and generalizable solubility prediction models. Table 1 summarizes the sample sizes and solubility distribution metrics for each dataset, while

Table 2 presents the molecular weight distribution metrics. Figure 2 visualizes the target variable distribution across datasets, and Figure 3 illustrates the molecular weight distribution ranges.

Table 1: Datasets size, distribution of solubilities or solvation free energy

| datasets | Num datapoints | Mean | Median | Standard deviation | Min | Max |
| --- | --- | --- | --- | --- | --- | --- |
| AqSolDB | 9982 | -2.89 | -2.62 | 2.37 | -13.17 | 2.14 |
| Arash | 8438 | -2.13 | -2.00 | 1.86 | -12.79 | 1.71 |
| ESOL | 1128 | -2.99 | -2.87 | 1.68 | -9.70 | 1.09 |
| FreeSolv | 642 | -3.80 | -3.53 | 3.85 | -25.47 | 3.43 |
| Samuel_acetone | 452 | -0.95 | -0.91 | 1.06 | -3.85 | 1.00 |
| Samuel_benzene | 464 | -1.17 | -1.06 | 1.08 | -3.93 | 0.97 |
| Samuel_ethanol | 695 | -1.32 | -1.33 | 1.12 | -3.96 | 0.97 |
| Samuel_water_narrow | 560 | -1.66 | -1.62 | 1.40 | -3.99 | 1.00 |
| Samuel_water_wide | 900 | -3.01 | -2.95 | 2.44 | -12.79 | 1.58 |

Table 2 Molecular weight (MW) distribution of the datasets

| datasets | Mean | Median | Standard deviation | Min | Max |
| --- | --- | --- | --- | --- | --- |
| AqSolDB | 266.67 | 228.68 | 184.17 | 9.01 | 5299.45 |
| Arash | 190.66 | 180.17 | 76.20 | 16.04 | 697.61 |
| ESOL | 203.94 | 182.18 | 102.74 | 16.04 | 780.95 |
| FreeSolv | 138.95 | 120.88 | 72.70 | 16.04 | 498.66 |
| Samuel_water_wide | 226.01 | 225.31 | 96.72 | 26.04 | 498.66 |
| Samuel_water_narrow | 197.97 | 197.15 | 85.53 | 26.04 | 495.40 |
| Samuel_acetone | 276.56 | 267.26 | 104.68 | 84.08 | 501.48 |
| Samuel_benzene | 234.74 | 235.62 | 86.57 | 87.12 | 500.60 |
| Samuel_ethanol | 270.46 | 256.09 | 100.54 | 69.07 | 504.63 |

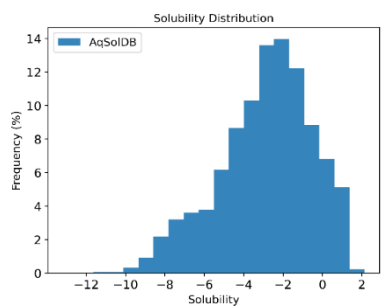

AqSolDB

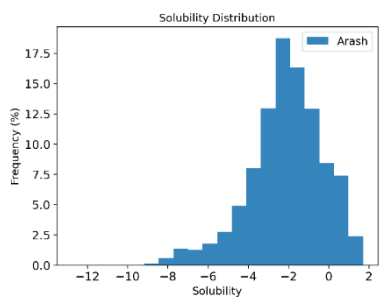

Arash

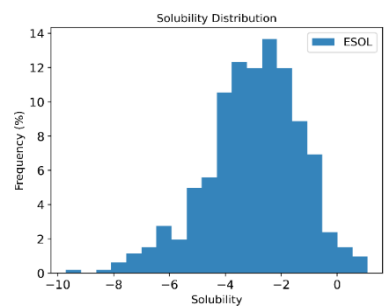

ESOL

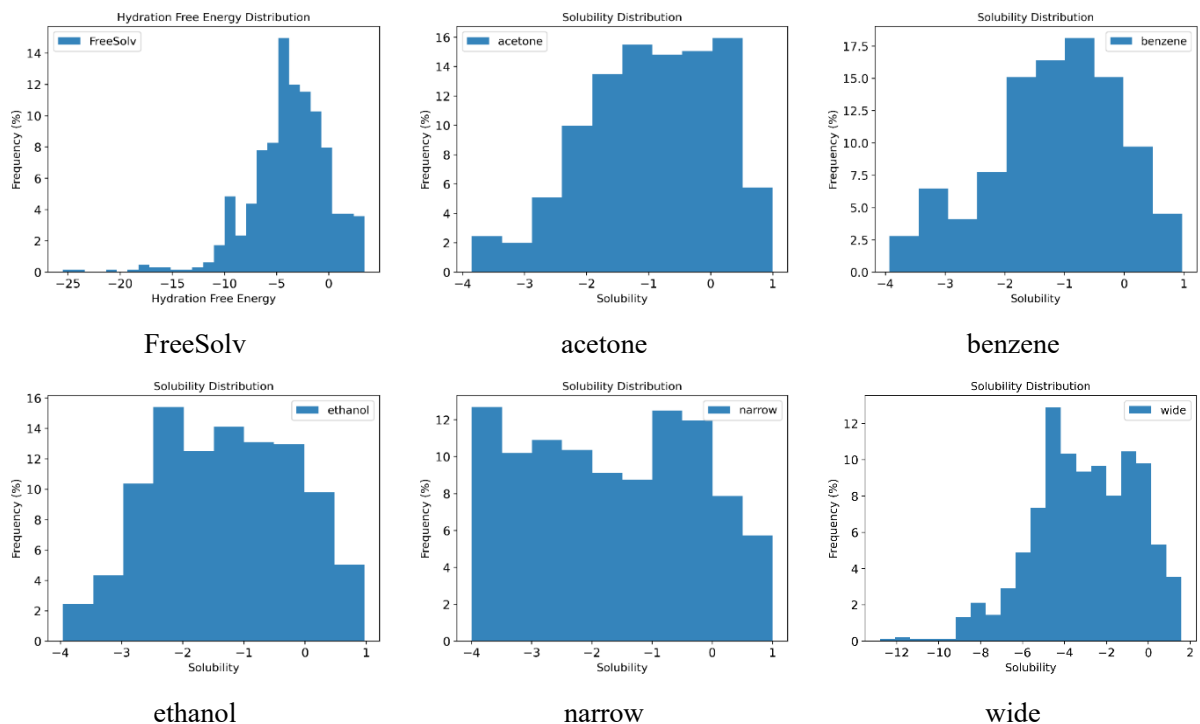

Figure 2 Histogram of solubility or solvation free energy

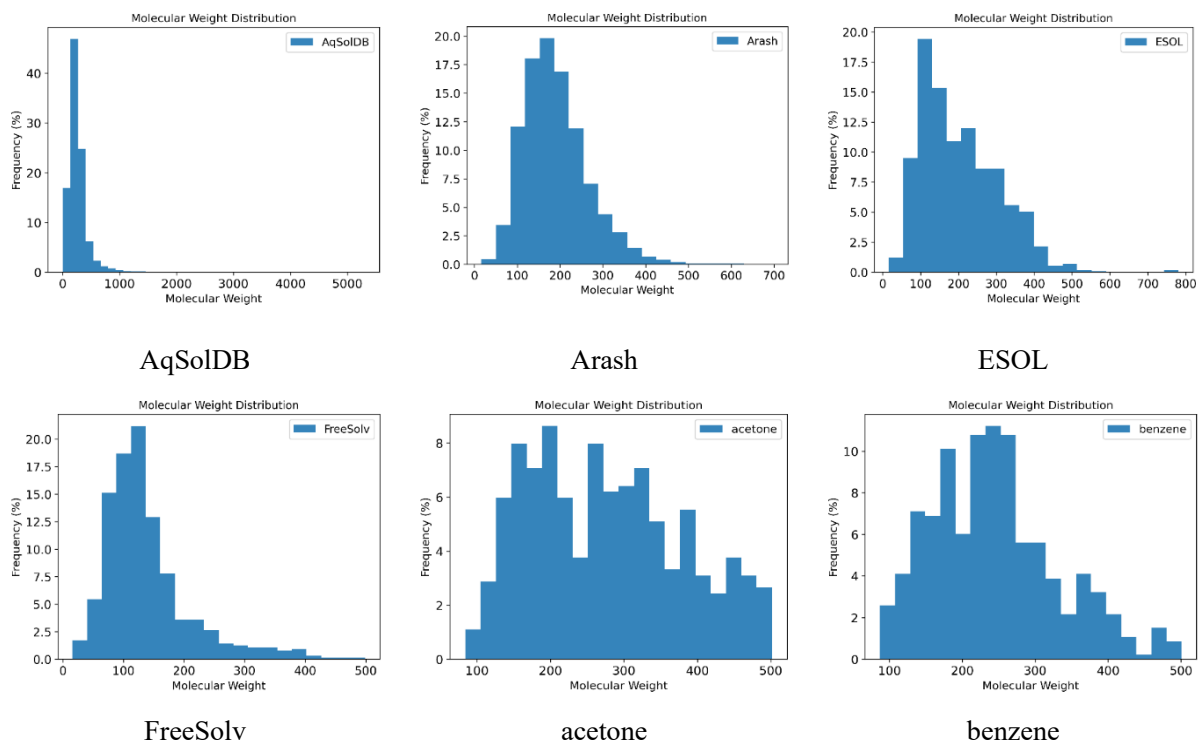

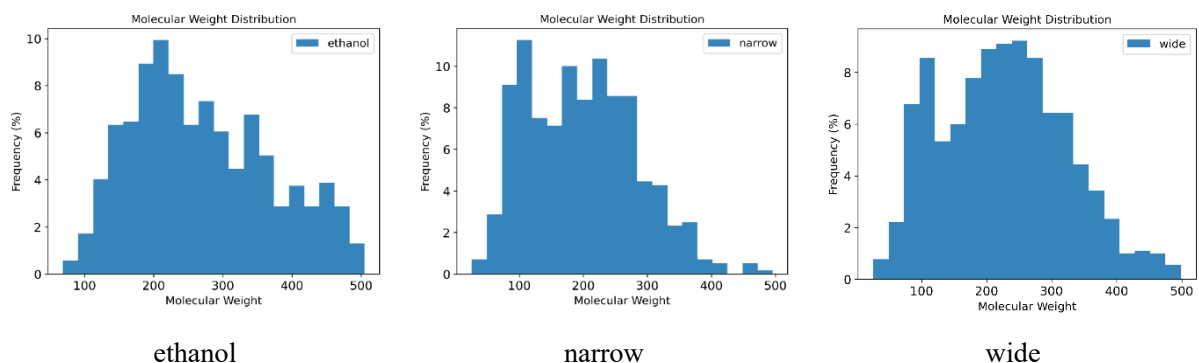

| ethanol | narrow | wide |

Figure 3: Histogram of molecular weight distribution for each dataset

Table 3 provides a detailed overview of the representation methods and their dimensions for different datasets. A "√" indicates that a specific representation method was applied, with the length of each fingerprint recorded accordingly. To further enhance predictive performance and model interpretability, we developed a multi-scale hybrid fingerprint (HF) molecular representation method that combines fragment fingerprints with a small set of physicochemical properties. We employed SMILES Pair Encoding (SPE)[45] and a functional group-based fragmentation approach. SPE first builds a vocabulary of frequently occurring SMILES substrings by learning from large chemical datasets such as ChEMBL, then tokenizes SMILES strings based on this vocabulary to facilitate deep learning model training. By introducing human-readable and chemically interpretable SMILES substrings as tokens, SPE improves upon traditional atom-level tokenization. The functional group-based fragmentation method, relying on domain knowledge, focuses on identifying and extracting key functional groups in compounds (e.g., heteroatoms, double bonds, triple bonds, acetal groups, and ring structures). This method leverages pattern matching and graph theory principles to decompose complex molecular structures into functionally meaningful fragments. The functional group-based approach complements SPE by adding domain-specific insights, enriching the molecular substructure information. For the global physicochemical properties (descriptors) in the multi-scale HF, we used descriptors generated by Mordred or those provided in the dataset. The descriptors and fragmentation information provide both global (macro-scale) and local (micro-scale) structural insights for molecular predictions, respectively. Additionally, compared to traditional binary (0-1) fingerprints, the descriptor and fragment frequency fingerprints offer a more compact molecular representation, further enhancing predictive performance.

Table 3 Molecular representations of the datasets

| Dataset | MACCS-Keys | Morgan-Fingerprints | Multi-scale HF |
| --- | --- | --- | --- |
| AqSolDB | √ (167bits) | √ (2048bits) | √ (68+243bits) |
| Arash | √ (167bist) | √ (2048bits) | √ (68+243bits) |
| ESOL | √ (167bits) | √ (2048bits) | √ (68+243bits) |
| FreeSolv | √ (167bits) | √ (2048bits) | √ (68+243bits) |
| Samuel_water_wide | √ (167bits) | √ (2048bits) | √ (68+243bits) |
| Samuel_water_narrow | √ (167bits) | √ (2048bits) | √ (68+243bits) |
| Samuel_acetone | √ (167bits) | √ (2048bits) | √ (41+139bits) |
| Samuel_benzene | √ (167bits) | √ (2048bits) | √ (41+139bits) |
| Samuel_ethanol | √ (167bits) | √ (2048bits) | √ (41+139bits) |

**4.2 Experimental Procedure and Comparison Methods**

In the aforementioned dataset, we compared the performance of RFM, linear regression, Gradient Boosting Tree[45], XGBoost[47], random forest[48], ResNet[49], and FT-Transformer[50] across different datasets, and demonstrated their prediction results using different molecular representations. For the widely used public datasets FreeSolv and ESOL, we benchmarked against more advanced prediction methods, such as GCN, GIN, D-MPNN, and state-of-the-art (SOTA) graph transformers. Subsequently, we analyzed the interpretability of RFM feature importance from both local and global perspectives. Finally, we examined the overfitting behavior of four kernel functions across the datasets.

**4.3 Experimental Results**

This section is divided into three parts: the first part validates the proposed RFM feature importance analysis method, demonstrating both its local (micro-scale) and global (macro-scale) multi-scale interpretability. The second part explores the extension of the RFM approach across different kernel functions, highlighting the model's performance in terms of overfitting resilience with each kernel type. The third part illustrates the significant performance advantages of RFM in QSPR modeling, where RFM combined with multi-scale HF even surpasses advanced GNN and graph transformer models in both accuracy and speed, underscoring its potential as a powerful new tool in molecular property prediction.

4.3.1 Feature Importance Analysis

For the feature importance analysis of RFM, we used the popular public benchmark datasets ESOL and FreeSolv as examples to demonstrate the correlation between the feature importance identified by RFM and those identified by PI and SHAP. Tables 4 and 5 list the number of common important features selected by RFM, PI, and SHAP across four different representations in the ESOL dataset. Out of the 167 MACCS Keys features, 8 of the top 10 features identified by RFM were consistent with those selected by PI, and 8 were consistent with SHAP. For the 2048 Morgan2 features, RFM shared 6 features with PI and 5 with SHAP. Among the 311 HF features, 6 of the top 10 features selected by RFM aligned with PI, and 5 were consistent with SHAP. Similarly, Table 5 presents the results for the top 20 most important features. The data in these tables show a high correlation between the feature importance identified by RFM and that identified by PI and SHAP, indicating that RFM has a high level of reliability in global feature analysis. Figure 4-5 provide a visual representation of the results in Tables 4 - 7.

Table 4: Feature Importance Analysis of RFM on the ESOL dataset (Top 10 Features)

| Methods/datasets | ESOL-RFM-10 | | |
|---|---|---|---|
| | MACCS(167bits) | Morgan2(2048bits) | HF(311bits) |
| PI | 8/10 | 6/10 | 6/10 |
| SHAP | 8/10 | 5/10 | 5/10 |

Table 5: Feature Importance Analysis of RFM on the ESOL dataset (Top 20 Features)

| Methods/datasets | ESOL-RFM-20 | | |
|---|---|---|---|
| | MACCS(167bits) | Morgan2(2048bits) | HF(311bits) |
| PI | 16/20 | 15/20 | 10/20 |
| SHAP | 14/20 | 13/20 | 11/20 |

Table 6: Feature Importance Analysis of RFM on the FreeSolv dataset (Top 10 Features)

| Methods/datasets | FreeSolv-RFM-10 | | |
|---|---|---|---|
| | MACCS(167bits) | Morgan2(2048bits) | HF(311bits) |
| PI | 3/10 | 4/10 | 9/10 |
| SHAP | 3/10 | 5/10 | 9/10 |

Table 7: Feature Importance Analysis of RFM on the FreeSolv dataset (Top 20 Features)

| Methods/datasets | FreeSolv-RFM-20 | | |
|---|---|---|---|
| | MACCS(167bits) | Morgan2(2048bits) | HF(311bits) |
| PI | 13/20 | 14/20 | 15/20 |

| | SHAP | 11/20 | 13/20 | 16/20 |
|---|---|---|---|---|

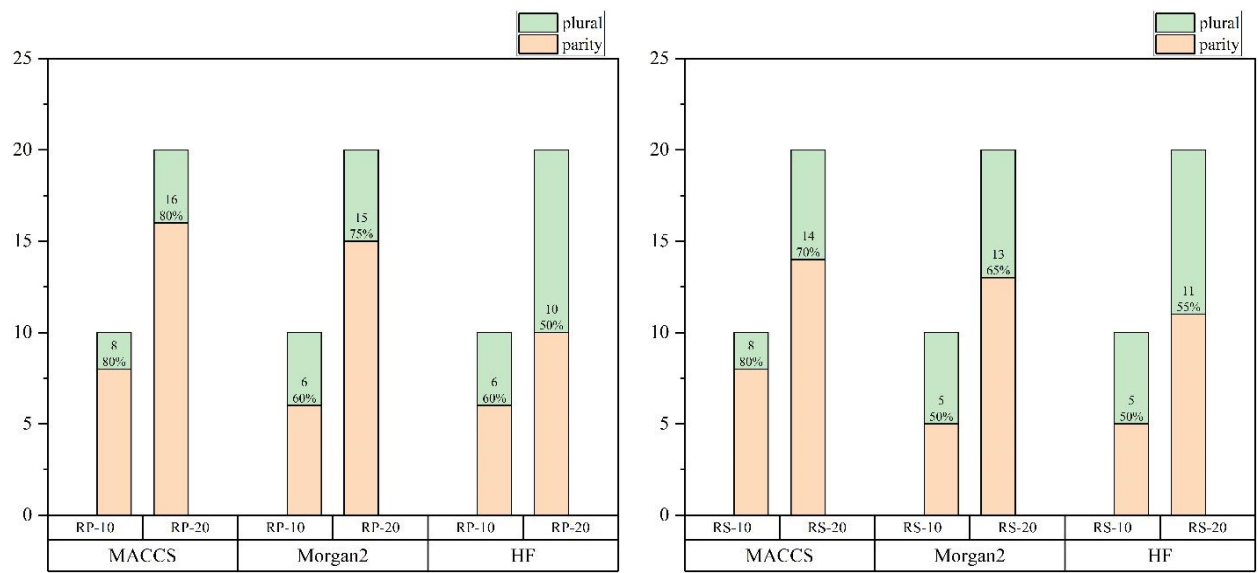

Figure 4: Correlation of Feature Analysis between RFM, PI, and SHAP on the ESOL dataset

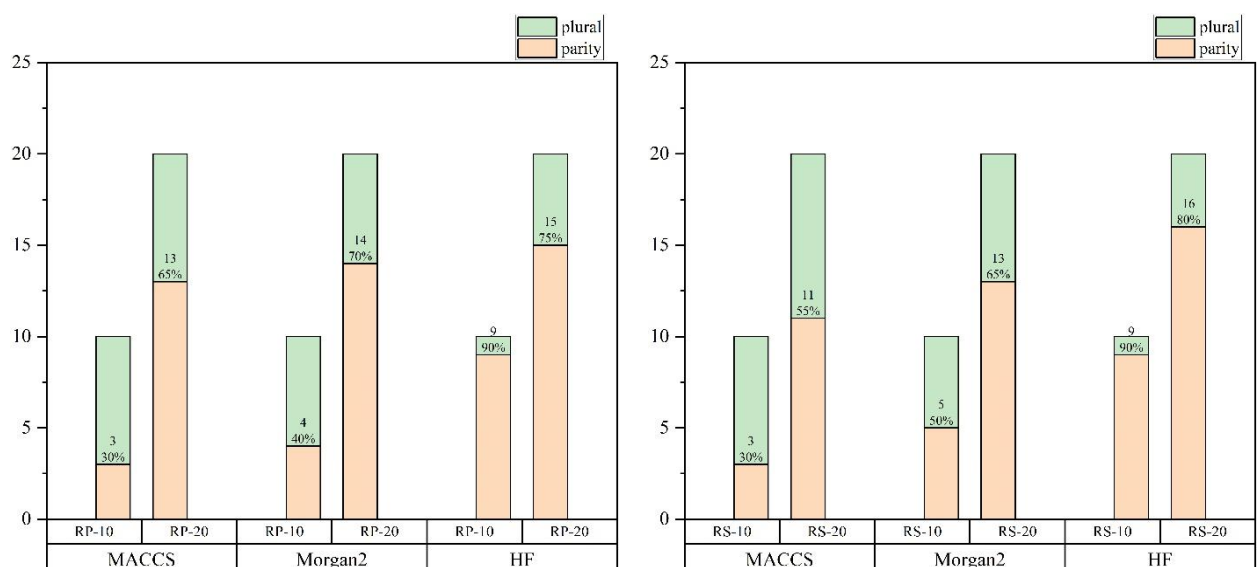

Figure 5: Correlation of Feature Analysis between RFM, PI, and SHAP on the FreeSolv dataset

Figures 4 and 5 respectively present the correlation analysis for the ESOL and FreeSolv datasets. In the figures, "plural" indicates the proportion of different features, and "parity" indicates the proportion of common features. "RP-10" and "RS-10" represent the number of common features between RFM and PI, and RFM and SHAP, respectively, among the top 10 important features. "RP-20" and "RS-20" indicate the number of common features among the top 20 important features.

In Table 8, we list the 20 most important features in the Arash dataset under the Morgan2 representation. The Morgan2 substructure diagrams in the table show the center atom in blue, aromatic atoms in yellow, alicyclic atoms in dark gray, and light gray structures representing features that influence atom connectivity, although these structures are not directly part of the fingerprint. Figure 6 provides the index and score ranking for these 20 features.

Table 8: Top 20 substructures with the highest scores among the 2048 features learned by RFM on the Arash dataset

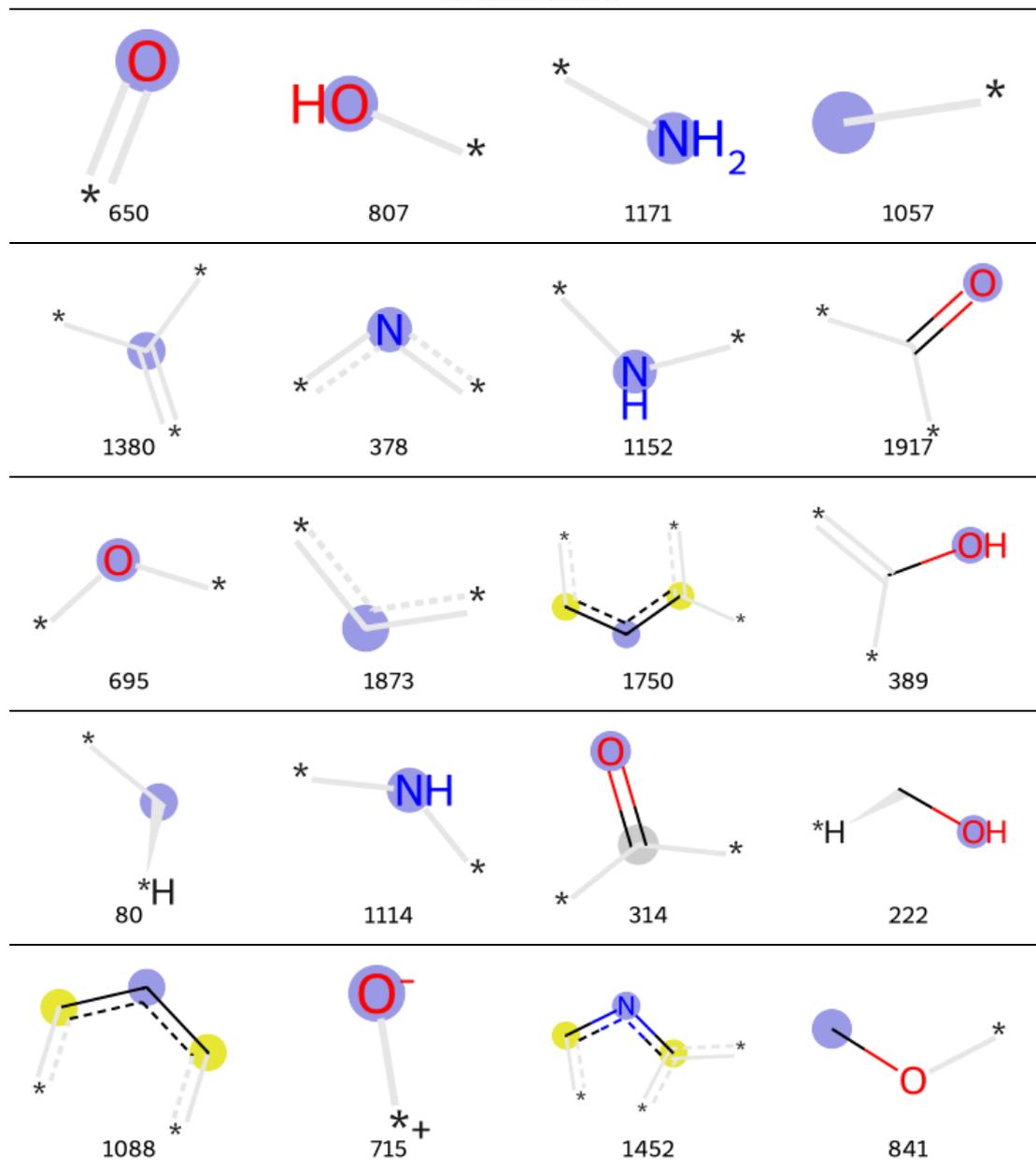

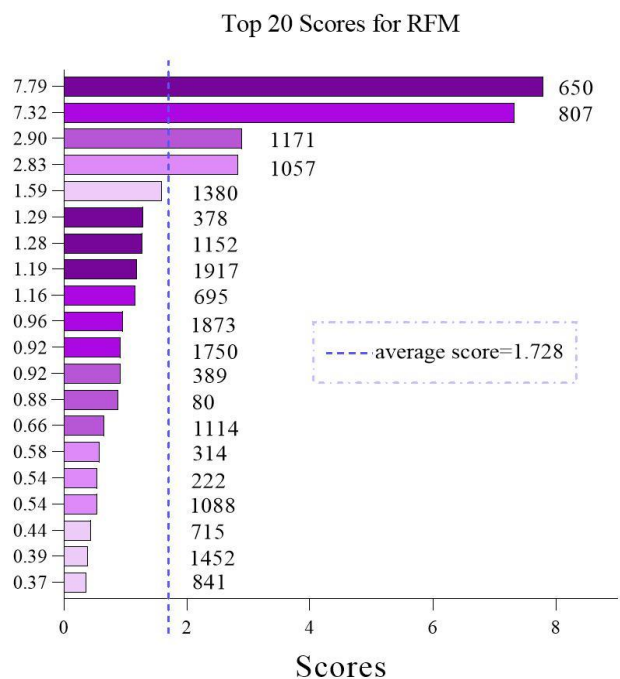

Figure 6: Top 20 Feature importance scores for the Arash dataset using the Morgan2 fingerprint.

The results above demonstrate the global interpretability of RFM. By comparing it with the feature importance analysis methods PI and SHAP, we confirmed the reliability of our proposed RFM global feature importance analysis. It is noteworthy that RFM's global interpretability is achieved through its local interpretability. RFM's local interpretability provides more detailed explanations for individual samples, allowing for an approximate analysis of whether certain features in a molecule have a positive or negative impact on its properties. In this study, we analyzed a soluble compound "29" and an insoluble compound "62." The solubility of "29" is LogS = -1.38, while the solubility of "62" is LogS = -5.259. Figure 7 shows the molecular structures of these two compounds, and Figure 8 illustrates the contribution of each feature within the molecule graphs to the overall graph property of "solubility."

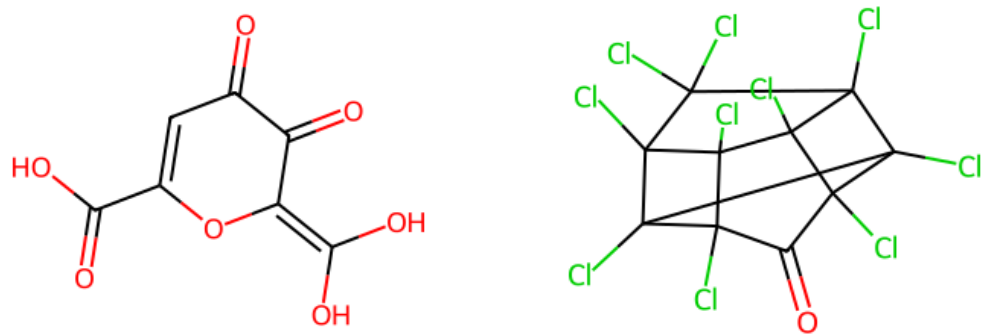

Figure 7: Molecular structure of compound "29" (Left) and compound "62" (Right)

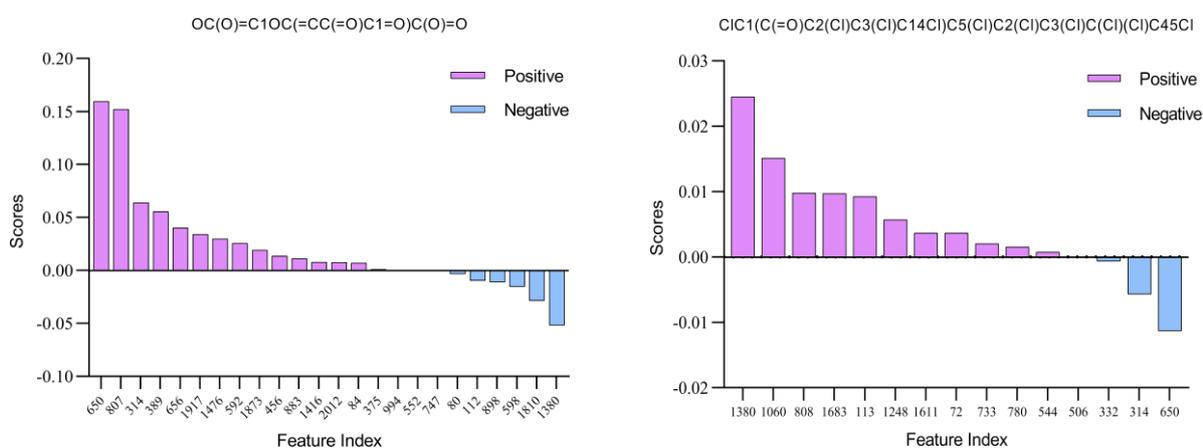

Figure 8: Features' contribution to the solubility of compound "29" (Left) and compound "62" (Right)

By observing Figure 8, we found that the same features, "650" and "314," contribute positively to the solubility in the molecule "29," while they have a negative impact in the molecule "62." Similarly, feature "1380" has a negative contribution in molecule "29," but it shows a positive contribution in molecule "62." Moreover, the scores of these features occupy a significant proportion of all features, indicating their crucial role in determining molecular solubility. Considering the solubility of molecules "26" and "29," where "26" is soluble in water and "29" is insoluble, we can infer that structures "605" and "314" make a molecule more water-soluble, whereas structure "1380" makes it less soluble in water.

These experimental results demonstrate that our proposed RFM feature importance analysis exhibits excellent interpretability at both the local and global levels, making it particularly suitable for the analysis of molecular features. We hope that this analytical tool can provide valuable assistance to biologists and chemists.

4.3.2 Overfitting Analysis of RFM with Different Kernel Functions

Next, the RFM model was compared across different kernel methods to find the best-performing kernel functions. The original RFM, incorporating AGOP with the Laplace kernel, demonstrated exceptional results. Further experiments incorporated AGOP into other kernel methods, including the Matern Kernel[51], Gaussian Kernel[52], and Rational Quadratic Kernel[53], with comparisons made across four different molecular representations in the AqSolDB dataset. Using MACCS representations in AqSolDB as an example, the performance was analyzed on both training and test sets, along with the eigenvalue spectra of the M matrix post-training, to evaluate overfitting levels for each kernel. Consistent regularization parameters were applied across all methods for a fair assessment.

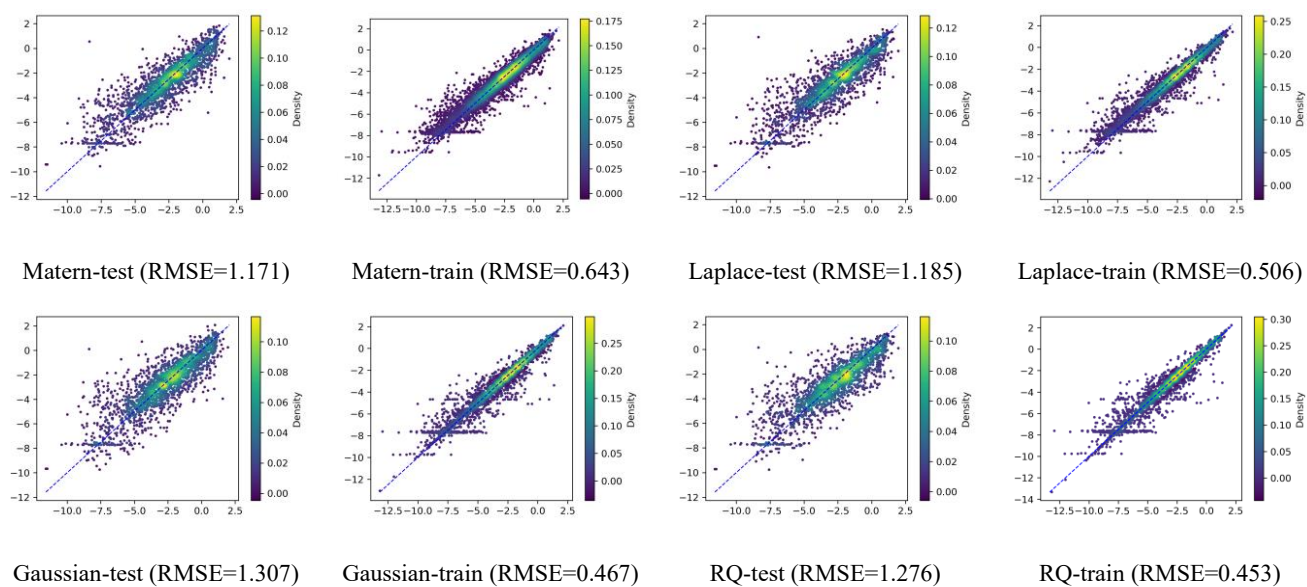

Matern-test (RMSE=1.171)　　Matern-train (RMSE=0.643)　　Laplace-test (RMSE=1.185)　　Laplace-train (RMSE=0.506)

Gaussian-test (RMSE=1.307)　　Gaussian-train (RMSE=0.467)　　RQ-test (RMSE=1.276)　　RQ-train (RMSE=0.453)

Figure 9: Prediction performance on the training and test sets of AqSolDB using the MACCS representation

Figure 9 illustrate the performance of the four kernels on the training and test sets. From these figures, we observe that the Matern and Laplace kernels exhibit similar levels of overfitting, followed by the Gaussian kernel, while the Rational Quadratic kernel demonstrates the most severe overfitting. Previous research on kernel function generalization indicates that the degree of overfitting is closely linked to the eigenvalue spectra of the empirical kernel matrix, where polynomial-like spectra exhibit milder overfitting compared to faster-declining exponential-like spectra, which may lead to catastrophic overfitting. Upon analyzing the eigenvalue spectrum of the empirical M matrix for RFM, we discovered similar patterns: the Rational Quadratic kernel's spectrum declines the fastest, followed by the Gaussian, while the Laplace and Matern kernels decline more gradually. This finding helps establish their generalization performance. Recent studies indicate that the Laplace kernel exhibits a polynomial spectrum in two-dimensional space without regularization, resulting in robust generalization even with some degree of overfitting. As an extension of the Laplace kernel, the Matern kernel inherits this advantage, demonstrating similar stability, which aligns with our experimental results. Figure 10 show the eigenvalue spectra of the different kernels.

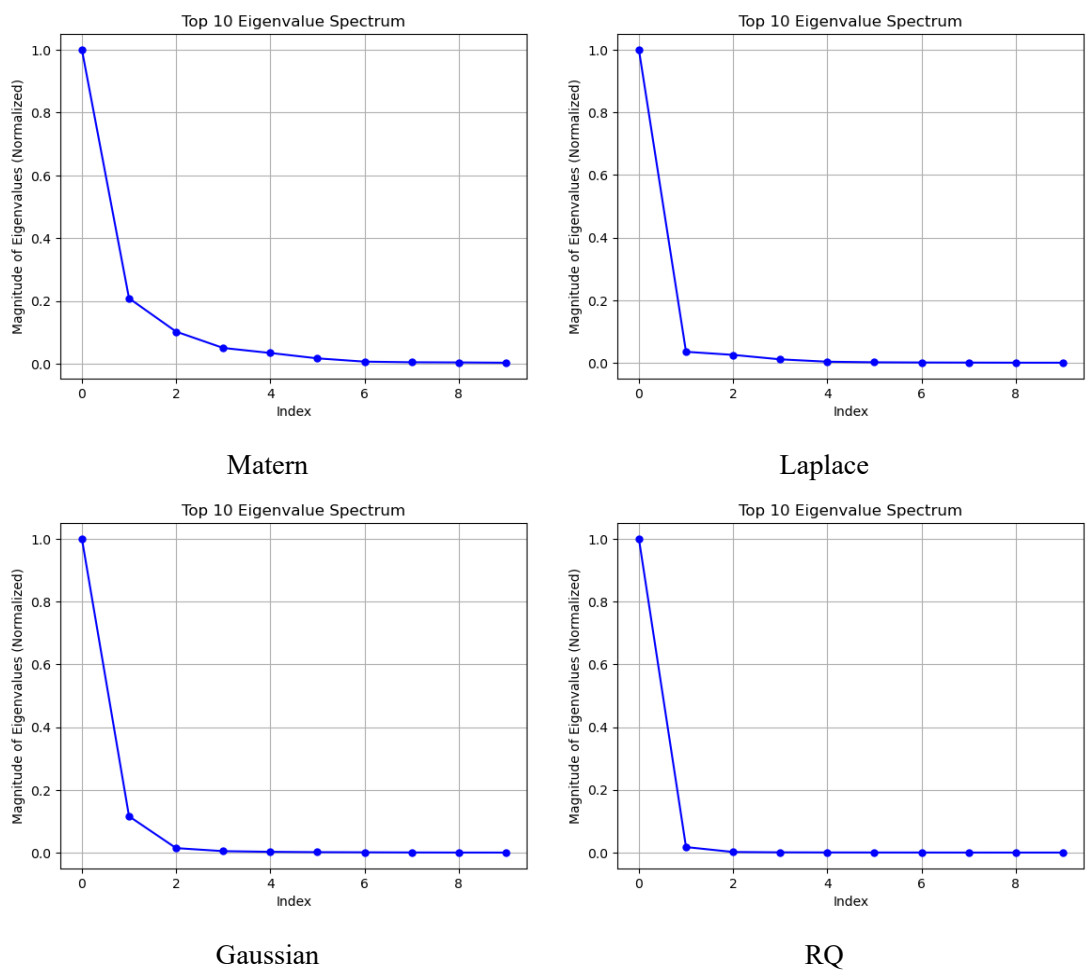

Figure 10: Eigenvalue spectrum of the M matrix for AqSolDB using the MACCS representation

### 4.3.3 Predictive Performance

This section first presents the predictive performance of each model, evaluated using $R^2$ and MSE as metrics. The results show that RFM performed exceptionally well across all datasets, balancing the predictive performance between the training and test sets, resulting in less overfitting compared to other methods. RFM achieved the best average R² and RMSE scores across nine datasets, demonstrating its significant advantage in handling tabular data. In practical applications, the coefficient of determination R² is typically used to compare the performance of different models. The closer the R² value is to 1, the more variance in the data the model can explain, indicating better performance. RMSE is a commonly used metric to measure the difference between the predicted and observed values; the smaller the RMSE, the closer the model's predictions are to the observed results, indicating superior model performance. The formulas for calculating these two metrics are as follows:

$$R^2 = 1 - \frac{\sum_{i=1}^{n}(y_i - \hat{y}_i)^2}{\sum_{i=1}^{n}(y_i - \bar{y}_i)^2} \tag{13}$$

$$RMSE = \sqrt{\frac{1}{n}\sum_{i=1}^{n}(y_i - \hat{y}_i)^2} \tag{14}$$

where $y_i, \hat{y}_i, \bar{y}_i$ are the observed, predicted and mean values respectively.

Table 9: R² using MACCS fingerprint representations and different ML methods

| Method \ Dataset | XGBoost | Random Forest | Gradient Boosting Tree | Linear Regression | ResNet | FT-Transformer | Laplace kernel-RFM |
|---|---|---|---|---|---|---|---|
| AqSolDB | <u>0.73</u> | 0.72 | <u>0.73</u> | 0.59 | 0.70 | 0.66 | **0.75** |
| Arash | <u>0.82</u> | 0.81 | <u>0.82</u> | 0.65 | 0.79 | 0.76 | **0.83** |
| ESOL | <u>0.81</u> | 0.79 | 0.80 | 0.37 | 0.79 | 0.75 | **0.84** |
| FreeSolv | 0.80 | 0.80 | <u>0.86</u> | - | **0.87** | **0.87** | 0.85 |
| water_wide | 0.78 | 0.73 | 0.81 | 0.66 | 0.80 | 0.74 | **0.82** |
| water_narrow | 0.66 | 0.64 | 0.63 | - | <u>0.68</u> | 0.50 | **0.69** |
| acetone | 0.26 | 0.24 | 0.25 | - | 0.13 | <u>0.28</u> | **0.32** |
| benzene | **0.23** | 0.19 | 0.15 | - | 0.16 | 0.11 | <u>0.20</u> |
| ethanol | 0.32 | <u>0.41</u> | 0.25 | - | 0.26 | 0.21 | **0.45** |
| Average R² | 0.60 | 0.59 | 0.58 | - | 0.58 | 0.54 | **0.64** |

Table 9 presents the Pearson correlation R² values obtained using MACCS Keys representation across nine different datasets. With the exception of the FreeSolv and Samuel_benzene datasets, where it slightly underperformed compared to other methods, RFM with the Laplace kernel achieved superior performance across all other datasets. RFM also significantly outperformed the other models in terms of the average R² value across the nine datasets. Typically, the R² coefficient ranges from [0, 1]. An R² of 1 indicates that the model perfectly predicts the data, while an R² of 0 indicates that the model perfectly predicts the data, while an R² of 0 means that the model fails to explain the variance in the data. The "-" in the Linear regression column indicates that the R² value was excessively low, rendering it meaningless for prediction. This suggests that the Linear regression model failed to capture any meaningful feature information and merely fitted random noise. The predictive scatter plots below, using the Arash dataset with MACCS Keys representation as an example, visually shows the predicted versus actual values on the test set, illustrating the predictive performance of each method. Results for the other datasets can be found in the Appendix.

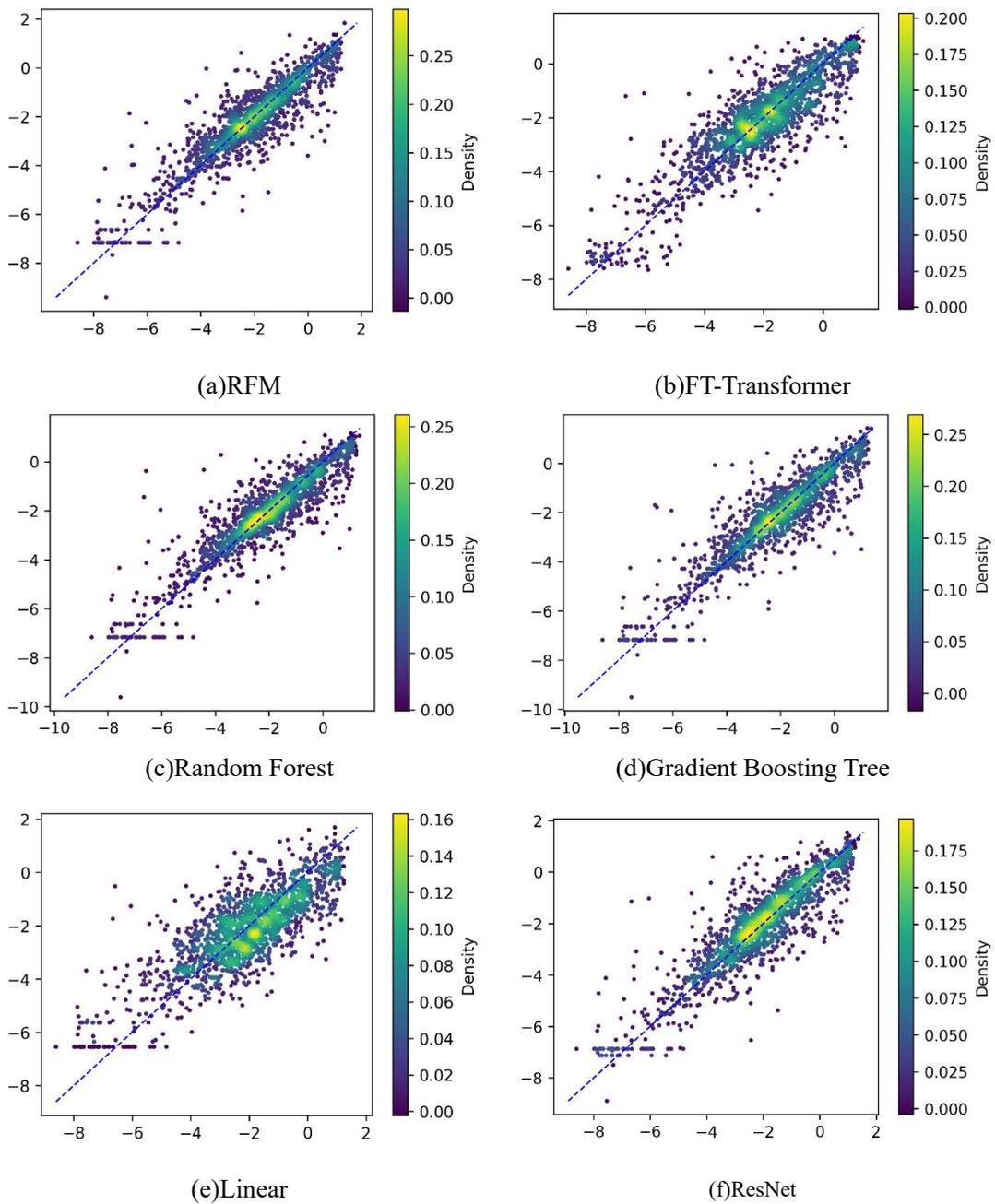

(a)RFM  (b)FT-Transformer

(c)Random Forest  (d)Gradient Boosting Tree

(e)Linear  (f)ResNet

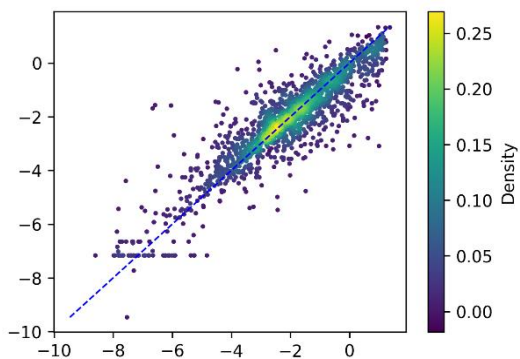

(g)XGBoost

Figure 11: Predictive scatterplots for each of the nine methods (Arash dataset, MACCS Keys representations)

Table 10: RMSE($\downarrow$) using MACCS fingerprint representations and different ML methods

| Method<br>Dataset | XGBoost | Random Forest | Gradient Boosting Tree | Linear Regression | ResNet | FT-Transformer | Laplace-kernel RFM |
|---|---|---|---|---|---|---|---|
| AqSolDB | <u>1.23</u> | 1.26 | 1.24 | 1.51 | 1.29 | 1.37 | **1.19** |
| Arash | <u>0.79</u> | <u>0.80</u> | <u>0.79</u> | 1.08 | 0.83 | 0.90 | **0.77** |
| ESOL | <u>0.73</u> | 0.76 | 0.74 | 1.32 | 0.76 | 0.83 | **0.67** |
| FreeSolv | 1.52 | 1.51 | 1.26 | - | <u>1.22</u> | **1.21** | 1.31 |
| water_wide | 1.10 | 1.21 | <u>1.02</u> | 1.37 | 1.05 | 1.21 | **1.00** |
| water_narrow | 0.85 | 0.88 | 0.88 | - | <u>0.82</u> | 1.04 | **0.81** |
| acetone | 0.94 | 0.95 | 0.95 | - | 1.02 | <u>0.93</u> | **0.90** |
| benzene | **0.87** | 0.90 | 0.92 | - | 0.91 | 0.94 | <u>0.89</u> |
| ethanol | 0.84 | <u>0.79</u> | 0.89 | 1.18 | 0.88 | 0.91 | **0.76** |
| Average RMSE | 0.99 | 1.01 | <u>0.97</u> | - | 0.98 | 1.04 | **0.92** |

Table 10 presents the Root mean squared error (MSE) results for the nine datasets using the MACCS Keys representation. Similar to the $R^2$ metric, RFM also demonstrated strong performance. In Table 10, the "-" indicates that the RMSE value for the linear regression method was excessively large, rendering it meaningless for prediction. By comparing the performance of DNNs models like the ResNet (CNN) and FT-Transformer with other methods in Tables 9 and 10, it can be observed that DNNs are less effective than traditional ML methods on these datasets. When the number of features and the sample size are relatively small, traditional ML methods perform better. All metrics were evaluated on the test sets, with the metrics for the training sets provided in the Appendix.

The following part of this section compares the predictive performance of three different molecular representations. The results indicate that multi-scale HF outperforms both MACCS and Morgan2 in terms of predictive ability. Compared to the fixed-length features of MACCS, HF provides more flexible and enriched substructural information by applying a correlation-based screening threshold, allowing better adaptability to the dataset's size and diversity. Additionally, unlike Morgan2 fingerprints, HF overcomes the issue of hash collisions, ensuring that each bit represents a specific chemical or biological feature, thereby reducing the information ambiguity caused by hash conflicts. From a machine learning perspective, the feature vectors generated by HF are more compact, containing not only local structural information but also implicitly capturing global information through descriptors. Since RFM relies on Kernel Ridge Regression (KRR), it may encounter issues like matrix singularity and overfitting when dealing with sparse vectors, especially high-dimensional sparse features. Compact vectors not only reduce computational overhead but also help kernel methods more effectively utilize local structural information within the compact representation. For complex ensemble models like XGBoost, Gradient Boosting Tree, and Random Forest, high-dimensional sparse vectors can also easily lead to overfitting. Using the AqSolDB, Arash, ESOL and FreeSolv datasets as examples, respectively, Tables 11-14 compare the predictive performance of traditional machine learning and deep learning methods under different molecular representations. Notably, our multi-scale HF outperformed both MACCS and Morgan2 fingerprints. By reviewing Tables 15 to 22 in the Appendix, we observed that the HF is generally applicable to various methods, with very few instances of failure.

Table 11: Results for AqSolDB using different ML and Representations methods (RMSE ↓ )

| Representations \Methods | MACCS(167bits) | Morgan2(2048bits) | HF(68+243bits) |
|---|---|---|---|
| RFM | 1.19 | 1.25 | **0.99** |
| XGBoost | 1.23 | 1.38 | **1.03** |
| Gradient Boosting Tree | 1.24 | 1.40 | **1.04** |
| Random Forest | 1.26 | 1.38 | **1.08** |
| Linear | **1.51** | 1.70 | **1.51** |
| FT-transformer | 1.37 | 1.60 | **1.14** |
| ResNet | 1.29 | 1.42 | **1.17** |

Table 12 Results for Arash dataset using different ML and Representations methods(RMSE ↓ )

| Representations \Methods | MACCS | Morgan2 | HF |
|---|---|---|---|
| RFM | 0.77 | 0.79 | **0.58** |
| XGBoost | 0.79 | 0.90 | **0.60** |
| Gradient Boosting Tree | 0.79 | 0.93 | **0.61** |
| Random Forest | 0.80 | 0.88 | **0.64** |
| Linear | 1.08 | 1.56 | **0.84** |
| FT-transformer | 0.90 | 1.06 | **0.70** |
| ResNet | 0.83 | 0.92 | **0.69** |

Table 13 Results for FreeSolv dataset using different ML and Representations methods(RMSE ↓ )

| Representations \Methods | MACCS(167bits) | Morgan2(2048bits) | HF(68+243bits) |
|---|---|---|---|
| RFM | 1.32 | 1.68 | **1.12** |
| XGBoost | 1.52 | 1.95 | **1.48** |
| Gradient Boosting Tree | **1.26** | 2.08 | 1.47 |
| Random Forest | **1.51** | 2.27 | 1.87 |
| Linear | - | - | **1.84** |
| FT-transformer | **1.21** | 2.85 | 1.46 |
| ResNet | **1.23** | 2.07 | 1.36 |

Table 14 Results for ESOL dataset using different ML and Representations methods(RMSE ↓ )

| Representations \Methods | MACCS(167bits) | Morgan2(2048bits) | HF(68+243bits) |
|---|---|---|---|
| RFM | 0.67 | 0.74 | **0.24** |
| XGBoost | 0.73 | 0.81 | **0.44** |
| Gradient Boosting Tree | 0.74 | 0.83 | **0.46** |
| Random Forest | 0.76 | 0.89 | **0.57** |
| Linear | 1.32 | - | **0.34** |
| FT-transformer | 0.83 | 1.28 | **0.31** |
| ResNet | 0.76 | 0.94 | **0.39** |

Tables 11-14 show the predictive performance of seven methods—including traditional ML methods such as XGBoost, Gradient Boosting Tree, Random Forest, Linear Regression, vs our kernel machine-based RFM, vs cutting-edge DNNs such as the Feature Tokenizer FT-Transformer, and ResNet—across three different molecular

representations. The results indicate that HF+RFM consistently outperforms MACCS Keys and Morgan2 in all scenarios. For instance, in Table 14, using RFM on the ESOL dataset, HF achieves a 43% improvement in predictive performance over MACCS Keys and a 50% improvement over Morgan2. For most datasets (Figures 12-15), the RFM-HF predictions show the tightest alignment with the ground truth, highlighting superior predictive accuracy.

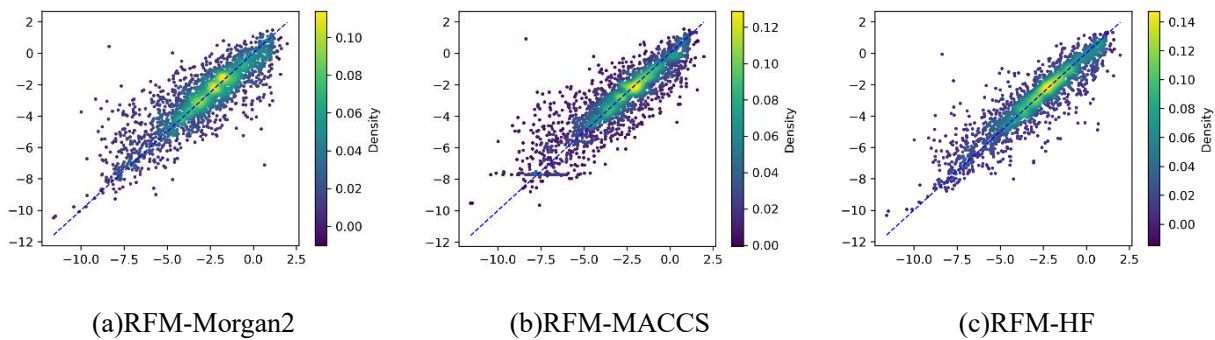

(a)RFM-Morgan2　　　　　　　(b)RFM-MACCS　　　　　　　(c)RFM-HF

Figure 12: Performance of RFM on the AqSolDB dataset using different chemical representations methods

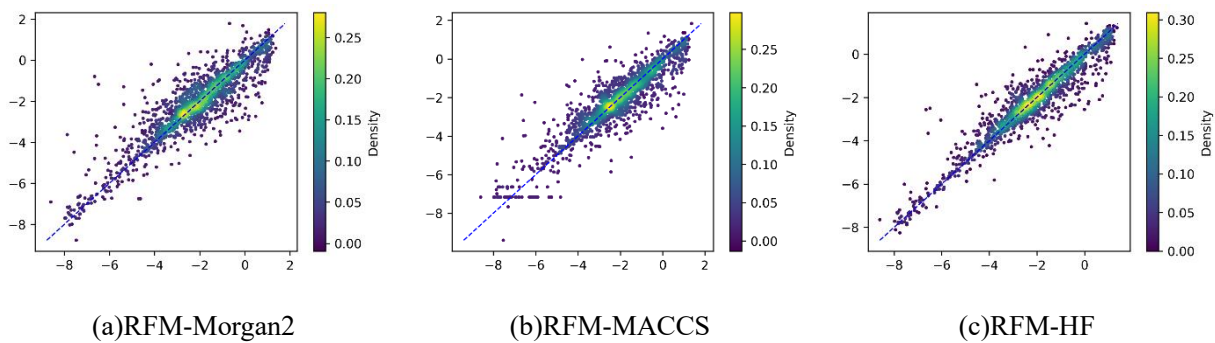

(a)RFM-Morgan2　　　　　　　(b)RFM-MACCS　　　　　　　(c)RFM-HF

Figure 13: Performance of RFM on the Arash dataset using different chemical representation methods

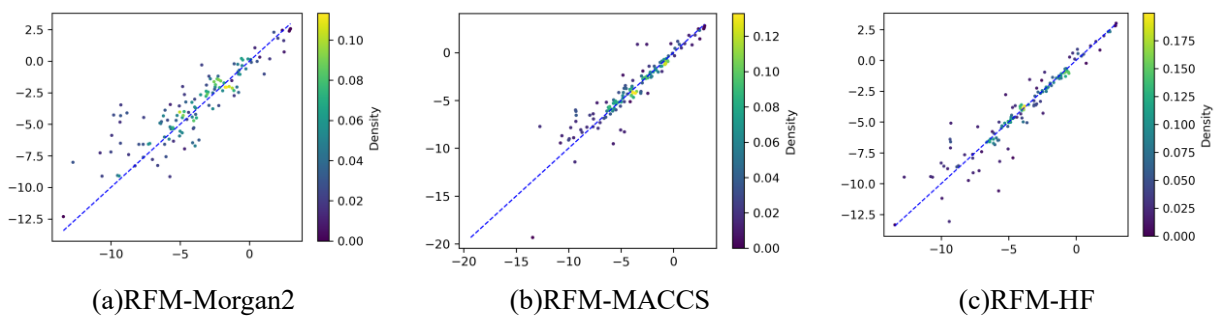

(a)RFM-Morgan2　　　　　　　(b)RFM-MACCS　　　　　　　(c)RFM-HF

Figure 14: Performance of RFM on the FreeSolv dataset using different chemical representation methods

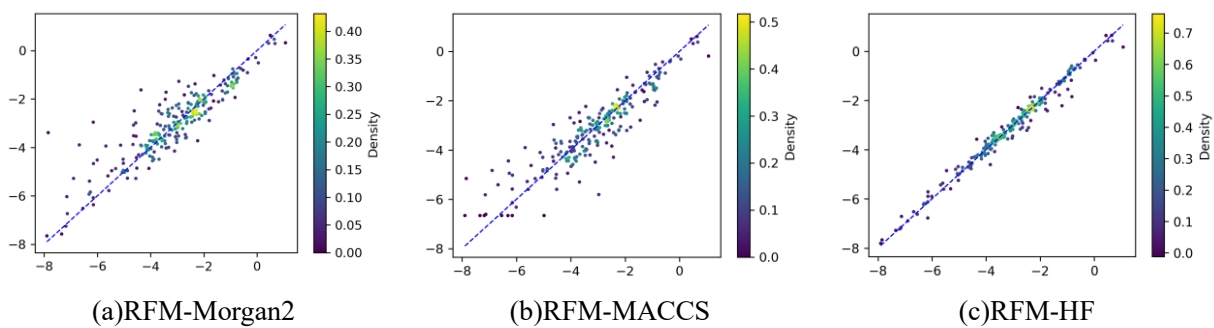

(a)RFM-Morgan2  (b)RFM-MACCS  (c)RFM-HF

Figure 15: Performance of RFM on the ESOL dataset using different chemical representation methods

For the popular benchmark ESOL and FreeSolv datasets, we conducted a comparative evaluation against several state-of-the-art (SOTA) models[54]. The results of ChemBFN[55], SPMM[56], ChemRL-GEM[57], and Uni-Mol[58] are referenced from their respective original publications. The RFM column in Table 15 reflects our own experimental results based on the superior Laplace kernel function and HF representations. Since the multi-scale HF representation achieved the best performance, we compared RFM with advanced graph neural networks, including the Multi-Scale Graph Convolutional Network (MGCN), SchNet (quantum-chemical deep tensor neural network), Graph Convolutional Network (GCN), Graph Isomorphism Network (GIN), and directed Message Passing Neural Network (D-MPNN). Additionally, GINE are molecular machine learning frameworks based on set representation learning, which implicitly represent topological structures using atomic and bond invariants, rather than explicitly defining them[54]. On both the ESOL and FreeSolv datasets, RFM-HF demonstrates exceptional predictive accuracy, outperforming all other models, including the above cutting-edge GNN architectures and Graph Transformer models. This superior performance highlights the effectiveness and efficiency of RFM-HF in the multi-scale learning of critical molecular features for enhanced QSPR modeling. In Table 15, the best results are highlighted in **bold**, and the second-best results are underlined. For metrics represented as $a \pm b$, $a$ denotes the average value from multiple experiments, while $b$ represents the corresponding standard deviation.

Table 15: RMSE (↓) results on two benchmark datasets for graph property (solubility and solvation free energy) prediction.

| Model \ Dataset | MGCN | SchNet | GCN | GIN | DMPNN | GINE | ChemBFN | SPMM | ChemRL-GEM | Uni-Mol | RFM-HF |
|---|---|---|---|---|---|---|---|---|---|---|---|
| ESOL | 1.27 ± 0.15 | 1.05 ± 0.06 | 1.43 ± 0.05 | 1.45 ± 0.02 | 0.98 ± 0.26 | 0.98 ± 0.10 | 0.884 | 0.810 | 0.798 | <u>0.788</u> | **0.24** |
| FreeSolv | 3.35 ± 0.01 | 3.22 ± 0.76 | 2.87 ± 0.14 | 2.76 ± 0.18 | 2.18 ± 0.91 | 2.92 ± 0.31 | <u>1.418</u> | 1.859 | 1.877 | 1.620 | **1.12** |

# Conclusion

Recursive Features Machines (RFM) is an innovative approach that approximate the deep feature representation capability of neural networks—by simply using traditional kernel learning models. In this study, we applied RFM to molecular property prediction modeling for the first time, using solubility as an example and conducting experiments on nine datasets with three types of representations: MACCS Keys, Morgan2 and our multi-scale HF. Compared to traditional ML methods like XGBoost, Random Forest, Gradient Boosting Tree, linear regression, and deep learning methods like ResNet, and FT-Transformer, the experimental results demonstrated that RFM outperforms all other methods in terms of predictive performance.

We proposed an RFM-based feature importance analysis method that offers both local and global interpretability, achieving global interpretability through local explanations. The effectiveness of our global interpretability method was validated through correlation analysis with two popular interpretability methods, PI and SHAP, while the effectiveness of local interpretability was demonstrated through chemical analysis of specific features within molecules. In addition to introducing AGOP into the Laplace kernel, we also applied it to Matern, Rational Quadratic, and Gaussian kernels. The experimental results showed that the Matern and Laplace kernels performed well, with the Gaussian kernel following closely behind. We also found that although regularization methods were introduced into the KRR model, this overfitting phenomenon still frequently occurs in some kernel methods. In contrast, in DNN models, over-parameterization seems to mitigate the bias-variance trade-off to some extent, maintaining good generalization despite the presence of overfitting noise. Many recent studies have also attempted to analyze this phenomenon in the relatively manageable environment of kernel regression [59].

In summary, we conducted an extensive comparison of different molecular representations and machine learning models for predicting molecular solubility and other molecular properties. The feature importance analysis method based on RFM allows for the analysis of data features from both local and global perspectives. This method is particularly well-suited for QSPR analysis modeling and contributes to the development of interpretable QSPR models, with the hope of advancing the fields of molecular design and drug discovery[59-60].

# Appendix

### RMSE - narrow

| Methods/Representations | MACCS(167bits) | Morgan2(2048bits) | HF(68+243bits) |
|---|---|---|---|
| RFM | 0.81 | 0.86 | **<span style="color:red">0.70</span>** |
| XGBoost | 0.85 | 0.97 | **0.78** |
| Gradient Boosting Tree | 0.88 | 1.01 | **0.80** |
| Random Forest | **0.88** | 1.02 | **0.88** |
| Linear | - | - | **1.21** |
| FT-transformer | 1.04 | - | **0.88** |
| ResNet | 0.82 | 0.91 | **0.81** |

### RMSE - wide

| Methods/Representations | MACCS(167bits) | Morgan2(2048bits) | HF(68+243bits) |
|---|---|---|---|
| RFM | 1.00 | 1.22 | **<span style="color:red">0.84</span>** |
| XGBoost | 1.10 | 1.43 | **0.91** |
| Gradient Boosting Tree | 1.02 | 1.37 | **0.94** |
| Random Forest | 1.21 | 1.43 | **0.96** |
| Linear | 1.37 | 1.22 | **1.21** |
| FT-transformer | 1.21 | 2.05 | **1.15** |
| ResNet | 1.05 | 1.42 | **0.93** |

### RMSE – FreeSolv

| Methods/Representations | MACCS(167bits) | Morgan2(2048bits) | HF(68+243bits) |
|---|---|---|---|
| RFM | 1.31 | 1.68 | **<span style="color:red">1.12</span>** |
| XGBoost | 1.52 | 1.95 | **1.48** |
| Gradient Boosting Tree | **1.26** | 2.08 | 1.47 |
| Random Forest | **1.51** | 2.27 | 1.87 |
| Linear | **1.32** | - | 1.84 |
| FT-transformer | **1.21** | 2.85 | 1.46 |
| ResNet | **1.23** | 2.07 | 1.36 |

### RMSE – ESOL

| Methods/Representations | MACCS(167bits) | Morgan2(2048bits) | HF(68+243bits) |
|---|---|---|---|
| RFM | 0.67 | 0.74 | **<span style="color:red">0.24</span>** |
| XGBoost | 0.73 | 0.81 | **0.44** |
| Gradient Boosting Tree | 0.74 | 0.83 | **0.46** |
| Random Forest | 0.76 | 0.89 | **0.57** |
| Linear | 1.32 | - | **0.34** |
| FT-transformer | 0.83 | 1.28 | **0.31** |
| ResNet | 0.76 | 0.94 | **0.39** |

| RMSE – Arash | | | |
|---|---|---|---|
| Methods/Representations | MACCS(167bits) | Morgan2(2048bits) | HF(68+243bits) |
| RFM | 0.77 | 0.79 | **<span style="color:red">0.58</span>** |
| XGBoost | 0.79 | 0.90 | **0.60** |
| Gradient Boosting Tree | 0.79 | 0.93 | **0.61** |
| Random Forest | 0.80 | 0.88 | **0.64** |
| Linear | 1.08 | 1.56 | **0.84** |
| FT-transformer | 0.90 | 1.06 | **0.70** |
| ResNet | 0.83 | 0.92 | **0.69** |

| RMSE – AqSolDB | | | |
|---|---|---|---|
| Methods/Representations | MACCS(167bits) | Morgan2(2048bits) | HF(68+243bits) |
| RFM | 1.19 | 1.25 | **<span style="color:red">0.99</span>** |
| XGBoost | 1.23 | 1.38 | **1.03** |
| Gradient Boosting Tree | 1.24 | 1.40 | **1.04** |
| Random Forest | 1.26 | 1.38 | **1.08** |
| Linear | **1.51** | 1.70 | **1.51** |
| FT-transformer | 1.37 | 1.60 | **1.14** |
| ResNet | 1.29 | 1.42 | **1.17** |